\documentclass[aps,prl,twocolumn,preprintnumbers,floatfix,superscriptaddress]{revtex4-1}
\usepackage{natbib,hyperref}
\hypersetup{colorlinks=true, citecolor=blue, urlcolor=blue, linkcolor=blue}
\usepackage[english]{babel}
\usepackage[utf8]{inputenc}
\usepackage{graphicx}
\usepackage[caption=false]{subfig}
\usepackage{amsmath}
\usepackage{amsfonts}
\usepackage{amssymb}
\usepackage{color,soul}
\setulcolor{red}
\usepackage{xcolor}

\begin{document}

\title{Relation between local density and density relaxation near glass transition in 
a glass forming binary mixture}
\author{D. C. Thakur}
\affiliation{School of Physical Sciences, 
Indian Institute of Technology Mandi,
Kamand, Himachal Pradesh 175005, India}
\author{Sandeep Kushawah}
\affiliation{School of Physical Sciences, 
Indian Institute of Technology Mandi,
Kamand, Himachal Pradesh 175005, India}
\author{Jalim Singh}
\affiliation{School of Physical Sciences, 
Indian Institute of Technology Mandi,
Kamand, Himachal Pradesh 175005, India}
\author{Anna Varughese}
\affiliation{School of Physical Sciences, 
Indian Institute of Technology Mandi,
Kamand, Himachal Pradesh 175005, India}
\author{Prasanth P. Jose}
\email{prasanth@iitmandi.ac.in}
\affiliation{School of Physical Sciences, 
Indian Institute of Technology Mandi,
Kamand, Himachal Pradesh 175005, India}


\begin{abstract} 
    Many  investigations shed light on various correlations between structure and 
       dynamics in supercooled liquids; however, a general relation between structure and
       dynamics remains elusive. This molecular dynamics simulation study identifies the interrelationship between 
       the growth of the highest 
    peak of the radial distribution function $g(r)$,   variation in the radial force from this 
    peak, and the slowdown of the density relaxation in the supercooled states of a model binary glass 
    former.   From the microscopic string-like motion 
    in supercooled liquids,
    we argue that the surface density on a spherical shell $\rho_{loc}$ around a 
    reference particle at the highest peak of $g(r)$ can represent the free volume available for 
    motion. We further show from these arguments and simulations
    that density relaxation time $\tau_\alpha$ and $\rho_{loc}$ are connected,
     $\tau_\alpha\propto\exp{(\rho_{loc}/(\rho_0-\rho_{loc}))}$, 
    where dynamics diverge at $\rho_0$. This relation is similar to the Vogel–Fulcher–Tammann 
    (VFT) relation in the supercooled liquids, thus giving insight into the structural origin of the VFT 
    as jamming of particles in a channel of density relaxation.
       \end{abstract}

\maketitle
Liquids undergo  glass transition or acquire solid-like rigidity with marginal structural
changes when compressed or cooled at a faster rate than structural relaxation.
These solid-like domains are due to the transient cages 
that resist density relaxation. Recent experiments on two-dimensional (2D) 
colloids show molecular-cages have higher local density \cite{j:boli}, whose 
size fluctuation facilitates the density relaxation \cite{j:past}. Therefore,
a microscopic theory of dense liquids can shed light on the relationship between
the structure of molecular cages and slowing down the density relaxation. 
A perturbation theory of dense Lennard-Jones (LJ) liquids by Weeks, Chandler, 
and Andersen (WCA) propose that repulsive interactions govern the structure 
of dense liquids \cite{j:wca}, which also suggest that the dynamics also 
follows the same \cite{j:chandler_1974}. A test of WCA theory in simulations  
of  glass forming binary mixture on a model (well-known for the study of glass
transition by Kob and 
Andersen (KA)  \cite{j:kobanderson2,j:stweb}) with (KA) and without  
attractive (KAWCA) interactions near the glass transition  
\cite{j:tarjus,j:tarjus1} show that the density relaxation dynamics is remarkably 
slower, in the supercooled state with attractive interactions at the number 
density $\rho=$1.2, while the density 
relaxation in KA and KAWCA models are nearly identical at the higher density  $\rho=$1.8. 
Studies also show that the relaxation dynamics become identical when the 
range of interaction includes the whole first coordination shell   
\cite{j:toxdyre}; besides, the attractive and repulsive type of  
interactions together form the fluctuating potential of the molecular 
cage, irrespective of the type; identical cage potential leads to the same 
density relaxation \cite{j:kawcapederson}. Therefore,  characterization 
of type, range, and depth of interactions of  molecular cages in the 
viscous liquids give insight into the relationship between dynamics and 
local structure.

A comparison of forces in KA and KAWCA models shows that the
difference between mean forces on particles are 
different when dynamics differ \cite{j:kawcalasse,j:chattoraj}. Also,  the 
configurational entropy obtained from the radial distribution function 
$g(r)$ \cite{tsl} can identify differences in the dynamics, thus suggesting that
$g(r)$ governs relaxation dynamics \cite{j:kawcasarika,j:ysingh}. Moreover,
insight from machine learning studies also supports the role of  
$g(r)$  \cite{j:landes}.  Interestingly, a recent study arrived at an
exponential relation between the relaxation dynamics of the system and 
order parameter derived from the relative angle between atoms in the 
first coordination shell that is similar to  Vogel–Fulcher–Tammann 
relation  \cite{j:tong}.    Therefore studies suggest that $g(r)$ at 
the first coordination shell has a decisive role in the relaxation 
dynamics of supercooled LJ liquids. Moreover, the schematic 
 mode-coupling theory uses the value of the structure factor at 
the first peak corresponding to the length scale of the nearest 
neighbor separation  \cite{j:bengtzelius,j:biroli}, to calculate 
density relaxation time.

Many studies support the relationship between local structure and 
relaxation dynamics. Notably, there is a relation between the short-time
$\beta$ relaxation (inside the cages) and $\alpha$ relaxation with 
both having identical temperature dependence \cite{j:karm},   the 
information theory also confirms the role of local parameters in defining 
global dynamics \cite{j:infthe}. Another study 
on gas-supercooled coexistence in polymers shows a correlation between 
density relaxation and particle distribution in the first 
coordination shell \cite{j:jalim1}.  Many recent studies also show 
that local potential affect dynamics \cite{j:locex} in 2D systems  
that is explained from dynamic facilitation \cite{j:dfac}; thus, 
suggest a relation between local structure and dynamics. The transient 
cage formation leads to an increase in local density \cite{j:boli} 
that reduces the available free volume for the molecules to relax. Thus
suggest that the free-volume theory \cite{j:grestrev} can address the 
relationship between structure and dynamics. In the free-volume theory 
of a hard-sphere system, the density relaxation 
$\tau_\alpha \propto \exp(B v^*_0/v^*_f)$, where $v^*_f=v^*-v^*_0$ 
is the difference between mean free-space and self-volume \cite{j:turnbull}.    
A rough estimate of free volume in soft-potentials is Voronoi volume 
\cite{j:starr1}, many studies look into  the detailed classification of
the various forms of free-volume, such as vibrationally accessible, 
hardcore free volume 
\cite{j:fvrev} \textit{etc.}.  This study deduces a relationship connecting 
the surface density $\rho_{loc}$  of a spherical shell with radius
$r^*$ at the highest peak of $g(r)\propto 1/v_f$, where $v_f$ is free-volume 
 in the channels of density relaxation and the density relaxation time
 $\tau_\alpha$,  from a systematic 
analysis of variation of $\tau_\alpha$  and $g(r^*)$ with temperature from
the analysis of the trajectories from simulations.   

\textit{Methods:}
Molecular dynamics simulations in this investigation 
use velocity Verlet algorithm \cite{b:allen} to integrate the equation 
of motion of $N=$ 8000 particles in a micro-canonical ensemble. 
The system simulated is the $4:1$ binary mixture  $A$ 
and $B$ defined in KA \cite{j:kobanderson2} and KAWCA \cite{j:tarjus1} 
models.    The KA potential as function of inter-particle separation 
$r$ reads:${V_{\alpha \beta}(r)}= 4 \epsilon_{\alpha \beta}{\left[\left({\sigma _{\alpha \beta} /{r}}\right)^{12} -{\left({\sigma _{\alpha \beta} / r} \right)^6} \right]+K_{\alpha\beta}}$, 
where parameters are $\alpha,\beta \in \{A,B\}, \epsilon_{AA}=1.0, 
\sigma_{AA}=1.0, \epsilon_{AB}=1.5, \sigma_{AB}=0.8, \epsilon_{BB}=0.5,$ 
and  $\sigma_{BB}=0.88$\cite{j:kobanderson2}; arbitrary constant $K_{\alpha\beta}$ ensures 
continuous potential at the cut-off. The  potential cut-offs are  
$2.5\sigma_{\alpha\beta}$ and $2^{1/6}\sigma_{\alpha\beta}$ respectively 
for KA and KWCA models. The density range of the simulation is from 
$\rho=$1.2 to 1.8 in a grid of $\delta\rho=$0.2; in each density,
temperatures form an uneven grid from high to low temperatures.  
The data presented in this study use Length, temperature, and time 
in units of $\sigma_{AA}$,  $\sigma_{AA} \sqrt{{m}/{\epsilon_{AA}}}$, 
and $\epsilon_{AA}/{k_B}$.  The initial configuration is prepared at 
high temperatures for all densities and quenched to the desired temperature.  
The duration of equilibration is 20 $\tau_\alpha$ or higher at 
all temperatures.

\begin{figure}
	\centering
	\includegraphics[width=6.0cm]{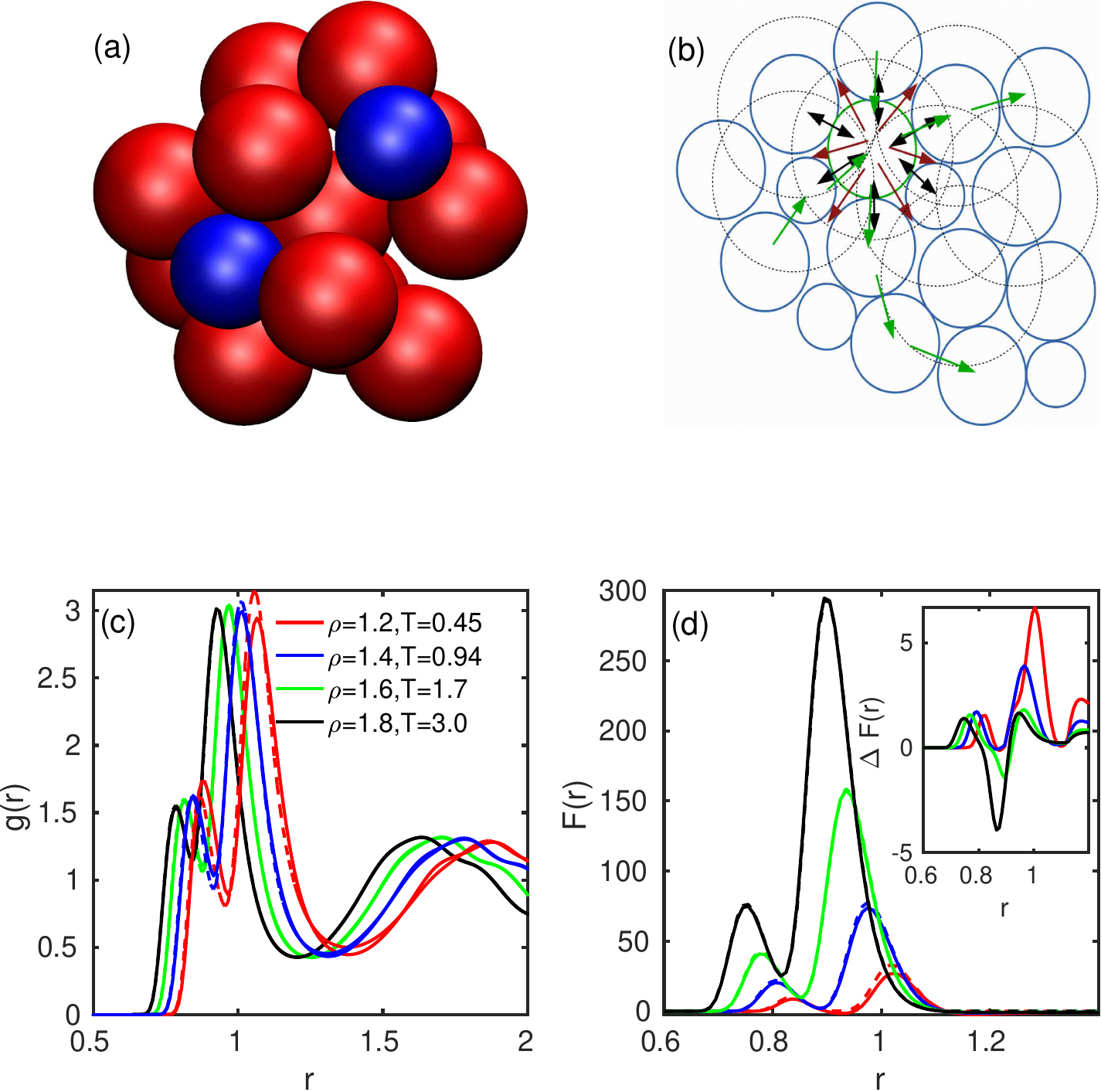}
	\caption{(a) Typical 3d of distribution 
of KA/KAWCA particles of a cage  at $\rho=$ 1.2 
($r_A \propto \sigma_{AA}$ and $r_B \propto (\sigma_{AB}+\sigma_{BB})/2$).
(b)The 2D schematic representation of (a)  with A  (big) and B (small). 
In (b): the green circle is the reference particle; the black double 
arrow is the preferred relaxation path; the red arrow is
unfavoured relaxation path; the green arrows are the string-like 
collective motion; dotted circles mark the typical peak of $g(r)$.
 (c) radial distribution functions $g(r)$ ; (d) corresponding variation in 
 the average radial force
$F(r)$ in supercooled at typical supercooled states; (d)(inset) shows difference
between $F(r)$ of KAWCA and KA models $\Delta F(r)$ at these state points.  }
	\label{f:f1}
\end{figure}
\begin{figure}
	\includegraphics[width=8.3cm]{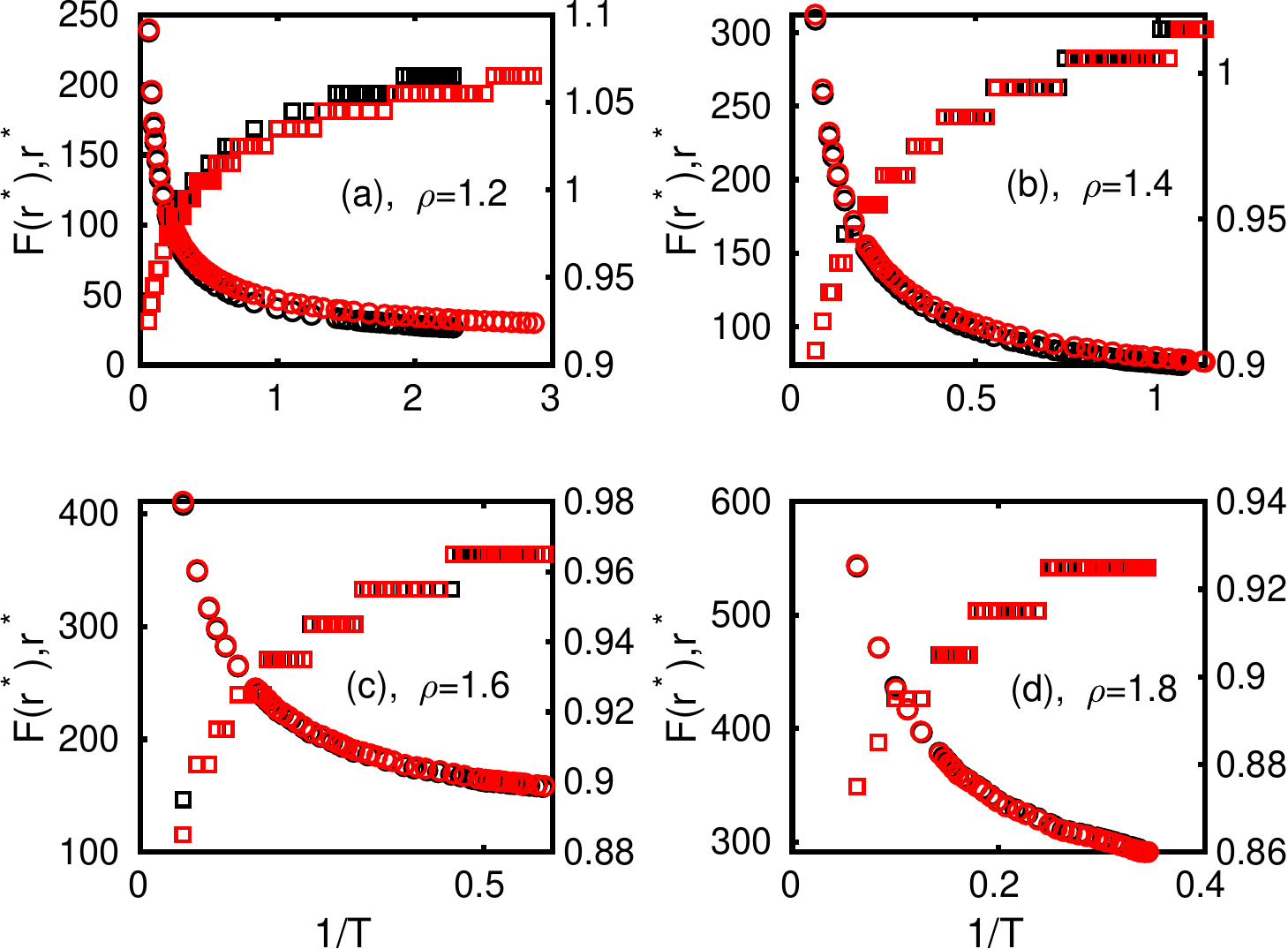}
\caption{Left axis and $\bigcirc$ show variation of $F(r^*)$ vs $1/T$.
    Right axis and $\square$ shows variation of $r^*$ (position of highest peak in $g(r)$)
vs $1/T$.   The color black and
red are for KA and  KAWCA, respectively. }
	\label{f:f2}
\end{figure}
\begin{figure}
	\centering
	\includegraphics[width=8.3cm]{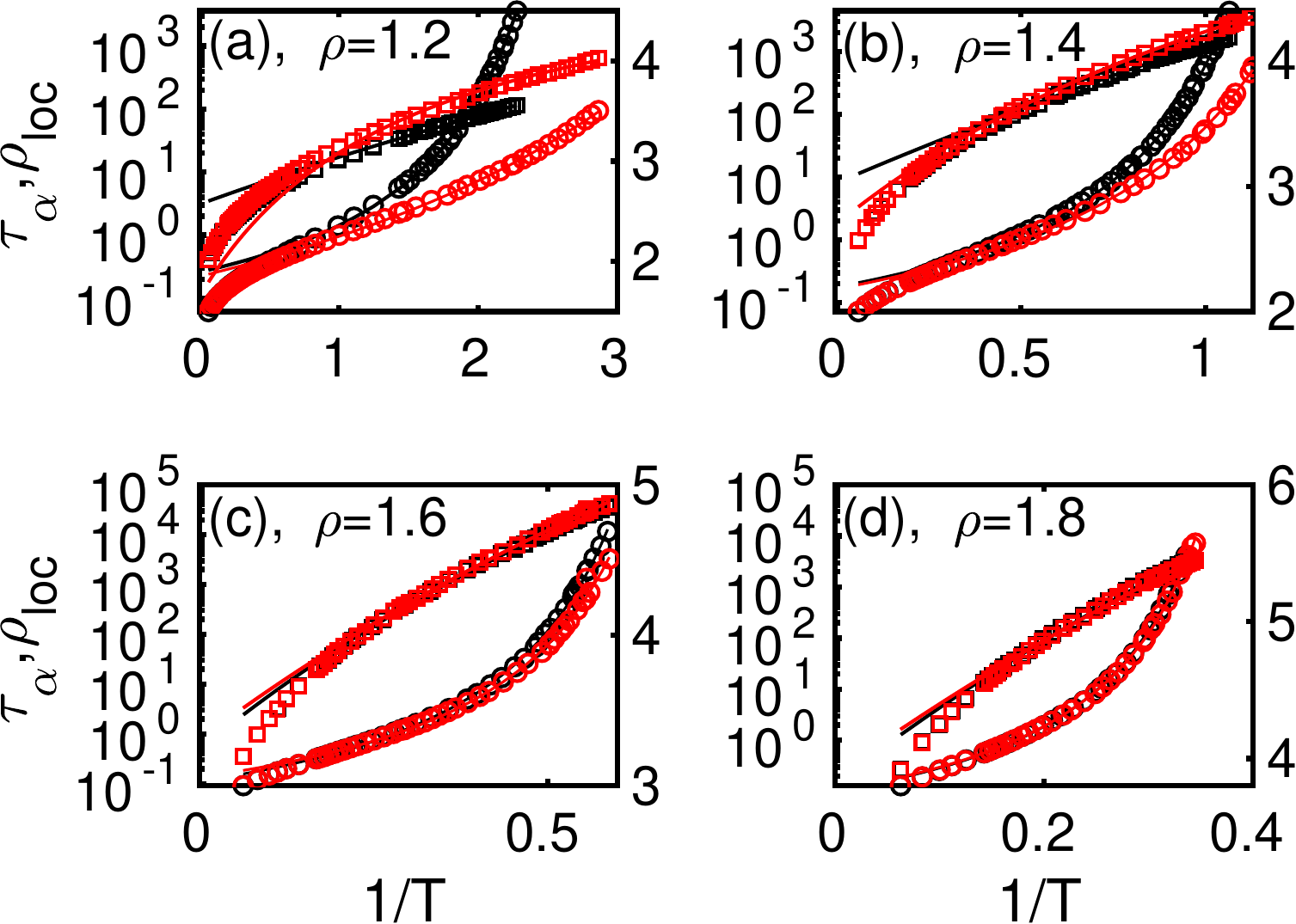}
	\caption{ $\tau_\alpha$ in semilog (left $Y$ scale) versus $1/T$ ,  
    by circles connected by VFT fit and $\rho_{loc}$ in linear 
scale(right $Y$ scale) versus $1/T$  as (squares), lines are 
Eq. \ref{e:tvft}, for KA (black) and KAWCA (red) models. }
	\label{f:f3}
\end{figure}

 We look at the first coordination shell in Figs. \ref{f:f1}(a) and (b) around a
 reference particle to arrive at a  relation between $g(r)$ and the available free volume, 
 where there is crowding of nearest neighbors in potential energy 
 minima. The  string-like motion  in the  supercooled 
liquids \cite{j:wk_prl,j:plimpton,j:aichele,j:yipstring} is a prominent microscopic 
mechanism of rearrangement of particles in the typical arrangement of particles 
in the schematic diagram given in Fig. \ref{f:f1}(b), 
where atoms move among nearly identical environments with a slight expansion of
the shell \cite{j:past}.  The string 
lengths enhance as temperature reduces in a supercooled liquid, \cite{j:wolstring2}.
The growth of the string lengths and the $\tau_\alpha$  with reduction of  temperature show similarity 
in simulation studies on the glass-forming supercooled polymers \cite{j:doug}.
Many studies identify that the strings that are part of string-like motion are cooperatively 
rearranging regions \cite{j:wolstring1, j:collstring,j:aksood} in theories of the 
glass-transition \cite{j:biroli}. Earlier studies show a preference for  
string-like motion at low $T$  is due to the high energy cost for the creation of  a hole  
\cite{j:lanlem,j:langer,j:polyflm}. The dominant type of particle motion 
that contribute to the relaxation 
are pair moves marked by the black arrows in Fig. \ref{f:f1}(b); many connecting
pairs lead to the typical string-like or ring-like collective motion (green arrows 
in Fig. \ref{f:f1}(b)) on a two-dimensional (2D) surface marked by the  $r^*$ 
which is common in the density relaxation of glasses \cite{j:swayamjyoti}.  The mean-field theory
for polymers \cite{b:flory,j:lanlem,j:langer} show that the free-energy barrier for 
string-like motion $E\propto v^{-1}_m$, where $v_m$ is  the available free-volume. 
The quasi-linear paths of strings are along the inter-penetrating  
2D surfaces marked as dotted circles in the one-dimensional (1D) projection 
in Fig. \ref{f:f1}(b). The number  density of particles along this path is proportional to
a peak height of the $g(r)$, which increases with the reduction of the temperature 
due to the unavailability of kinetic energy that prevents particles from
escaping from the potential energy minima and thus forms a cage.   
This cage exerts a fluctuating force on
any reference particle whose mean at position $r^*$ quantifies the barrier for 
a confined particle.  The mean  force  exerted on a reference particle  by 
density  at $r$ in the radial direction is 
$F(r)= \frac{1}{N\rho}\sum_{\alpha\beta} |F_{\alpha\beta}(r) \hat r|$, where  
$F_{\alpha\beta}(r)\hat r=\mathbf{f}_{\alpha\beta}(r)n_{\alpha\beta}(r)$ and, 
$\mathbf{f}_{\alpha\beta}(r)= - \frac{\partial V_{\alpha\beta}(r)}{\partial r} \hat r$, 
 where $n_{\alpha\beta}$ is 
unnormalized pair density. 
  $F(r)$ at the highest peak in Fig. \ref{f:f1}(d)  at $\rho=$1.2, differ for KAWCA and KA: plot of the difference  
in forces $\Delta F(r)$ in these models is in the inset of fig.\ref{f:f1}(d).  
 The $\Delta F(r)$  reduces at high densities $\rho=$1.8.  Next these growths
 in $g(r^*)$ and $F(r^*)$ in KA and KAWCA models
 are compared with
the density relaxation time. We get the estimate  of the density relaxation time $\tau_\alpha$  by integrating the  
incoherent intermediate scattering function $F_s(k,t)$\cite{tsl}, 
$\tau_\alpha=\int_0^\infty F_s(k,t) dt$  
at $k=$7.28, where $k$ is the mean wavenumber at the first peak of $S(k)$; see Fig. \ref{f:f3}(a-d).  The 
difference in mean force on particles in these models shows a difference
in relaxation in earlier studies \cite{j:kawcalasse,j:chattoraj}. 
The increase of  $\rho_{loc}$ results in a corresponding
exponential increase of $\tau_\alpha$ as the temperature 
in fig \ref{f:f3}, which is due to an increase in the jamming of particles 
in the relaxation channel marked with black arrows in fig.\ref{f:f1}(b).
Studies of the structure near jamming transition in repulsive 3D colloidal systems show 
that  $g(r)$ grows systematically near jamming 
\cite{j:colljam}. The signatures of jamming appear in the 
force distribution of bio-polymers \cite{j:ppjjam} predominantly due to attractive interaction.  

\begin{figure}
	\centering
	\includegraphics[width=8.3cm]{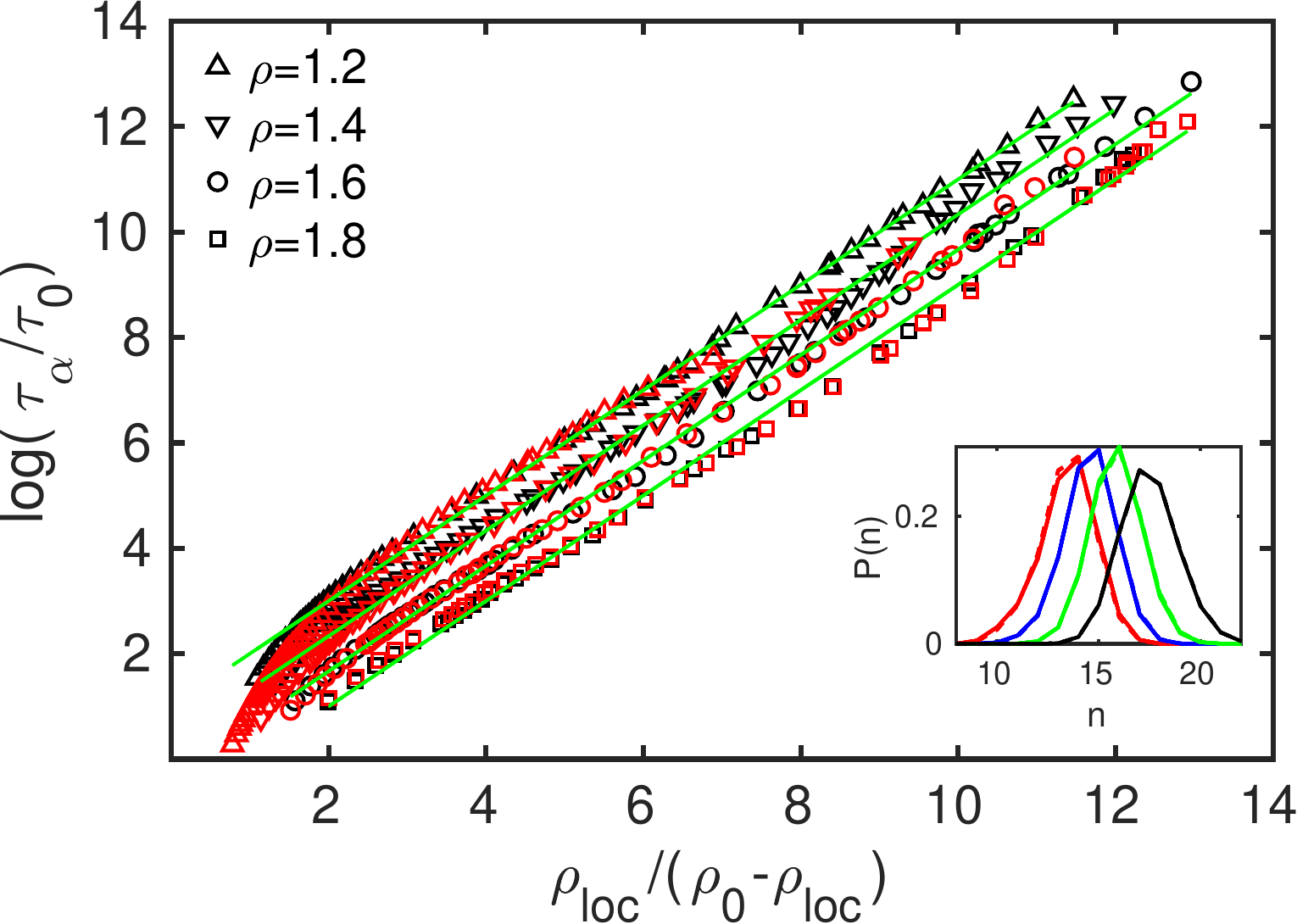}
	\caption{The collapsed plot of the logarithm of scaled 
    relaxation time versus relative variation of the local density
    $\rho_{loc}$ for KA (black) and KAWCA(red) models(shifted for clarity).
    Inset shows nearest 
    neighbor distribution $P(n)$ with in radius 1.4$\sigma_{AA}$ for all densities in multiple
supercooled states which overlap for a constant $\rho$. In the inset, the density
increase from left to right for KA($-$) KAWCA($--$).}
	\label{f:f4}
\end{figure}


The free-volume change in the spherical shell  $N_m /(\rho_0-\rho_{loc}) $,
where $N_{m}$ is the mean shell occupation in the supercooled state and
$\rho_0$ is the maximum value $\rho_{loc}$ attains
by decrease of $T$ for a particular $V$, beyond
which relaxation time diverges.  Note that in the supercooled state, 
nearest neighbor distribution $P(n)$ overlaps for many temperatures for
both models at a particular $\rho$; see the inset of 
Fig. \ref{f:f4}; implying $N_m$ is constant in the supercooled state.
There is negligible  variation in the number of particles in a spherical shell of 
infinitesimal thickness with $r^*\simeq$ constant 
(fig. \ref{f:f4}(inset)). This free volume in a shell with radius $r^*$  gives an estimate of
the overall free volume, which assists the density relaxation.
Then the relaxation time from 
free-volume theory is modified with $v_0=N_m/ \rho_0$ as a function of 
$\rho_{loc}$ reads
 \begin{equation}
    \tau_\alpha=\tau_\rho e^{\left( \frac{Bv_0}{v-v_0}\right)}\simeq   \tau_\rho e^{\left(\frac{B\rho_{loc}}{\rho_0-\rho_{loc}}\right)}.
	\label{eq:rht}
\end{equation}

A fit of this relation in  KA and  KAWCA models    for 
 Fig. \ref{f:f4} show $B\sim$ 1 \cite{j:doolittle}.
 The  deviation of the simulation data from Eq.\ref{eq:rht} is due to
 the  non-monotonous increase of $r^*$ in Fig. \ref{f:f2}  due to the
 formation of local structures that affect the shape of the cages.
 Moreover, the assumption of a perfectly
 spherical barrier of the cage marked by circles in \ref{f:f1}(b) is 
 the nearest approximation of the collective potential barrier of a molecular 
 cage. Note that the relaxation time fluctuates around a straight line in 
 Fig. \ref{f:f4}, showing 
 that this relation holds reasonably well.
 Also, a system of linear supercooled flexible polymers obeys 
 an equation which is similar to Eq. \ref{eq:rht} \cite{j:jalim}.
The eq. \ref{eq:rht}  is  similar to the Vogel–Fulcher–Tammann (VFT) 
relation $\tau_\alpha=\tau_T \exp(A/(T-T_0))$, where $A$ and $T_0$ are
constants.
Earlier experimental studies  show that the free volume is a function of 
temperature  $v^*_f=C+k(T-T_g)$ with constants $k$ and $C$, where $C$ is the excess 
free-volume at $T_g$  \cite{j:doolittle,j:mlwilliams}.  
With the new definition of free-volume of a pair of neighbors,
$v_f \propto(T-T_0)$ as $\rho_{loc}$ increase 
till $T_0$.  Many studies use an alternative representation 
of the free volume as a  function of the Debye-Waller factor 
$\langle u^2\rangle$ \cite{j:hall,j:starr1,j:simmons,j:larini},  
which shows an exponential increase of density relaxation time with 
$\langle u^2\rangle$ , which is 
comparable to Eq.\ref{eq:rht}  \cite{j:jalim}; thus suggest a 
relation between $\rho_{loc}$ and $\langle u^2\rangle$.  A study 
of string-like motion in the  polymer models show $\langle u^2\rangle$ 
vanish at $T_0$ \cite{j:strigletdougles}.    Thus equating the 
Eq. \ref{eq:rht} and VFT expression yields a relation between 
$\rho_{loc}$ and temperature difference reads 
\begin{equation}
    \rho_{loc}=\rho_0 \left[\frac{k (T-T_0)-A}{(k-1)(T-T_0)-A} \right],
	\label{e:tvft}
\end{equation}
where constant $k=\ln{(\tau_\rho/\tau_T)}$.
 In the supercooled liquids Eq. \ref{e:tvft} 
is valid; that is evident from the plot of Eq. \ref{e:tvft} in
Fig. \ref{f:f3}, which shows that the VFT relation is related to 
jamming of particles due to density enhancement, below this temperature
dynamics diverge significantly, arresting the density relaxation.

  In summary, these studies show a direct correlation between the local 
density and  the radial forces,  and the density relaxation 
dynamics. These results, when combined with 
string-like excitation \cite{j:wk_prl,j:plimpton,j:aichele,j:yipstring} 
and Flory's mean-field theory of polymer relaxation based on the 
free-volume \cite{b:flory,j:lanlem,j:langer} connects surface density at the peak
of the $g(r^*)$ in the first-coordination shell and the density relaxation 
time $\tau_\alpha$
in Eq.\ref{eq:rht}. The VFT-like structure of the Eq. \ref{eq:rht} is in 
agreement with a similar 
relation for an order parameter derived among the angle between particles in 
the first coordination shell \cite{j:tong},  which is valid in 2D as well. 
Therefore, the results of this study provide a model connecting local 
density and density relaxation in nonassociated liquids \cite{chandler}. 
 
\begin{acknowledgments}
    We thank the HPC facility at IIT Mandi for computational facilities. 
    The School of Physical Sciences is bifurcation
    of the former School of Basic Sciences at IIT Mandi, where DCT, 
    JS, and AV are affiliated
    during this work.
\end{acknowledgments}

%

\end{document}